\documentclass[a4paper]{article}
\usepackage{INTERSPEECH2021}
\usepackage{subfig}
\usepackage[utf8]{inputenc}

\title{Articulatory Coordination for Speech Motor Tracking in Huntington Disease}
\name{Matthew Perez$^1$, Amrit Romana$^1$, Angela Roberts$^2$, Noelle Carlozzi$^3$,  Jennifer Ann Miner$^3$, \\Praveen Dayalu$^4$, Emily Mower Provost$^1$}
\address{
  $^1$Computer Science and Engineering, University of Michigan, Ann Arbor, Michigan, USA\\
  $^2$Communication Sciences and Disorders, Northwestern University, Evanston, Illinois, USA\\
  $^3$Physical Medicine \& Rehabilitation, University of Michigan, Ann Arbor, MI \\
  $^4$Michigan Medicine, University of Michigan, Ann Arbor, MI}
\email{mkperez@umich.edu, aromana@umich.edu, angela.roberts@northwestern.edu, carlozzi@med.umich.edu, jenminer@med.umich.edu, pravd@med.umich.edu, emilykmp@umich.edu}

\begin{document}

\maketitle

\begin{abstract}



Huntington Disease (HD) is a progressive disorder which often manifests in motor impairment.
Motor severity (captured via motor score) is a key component in assessing overall HD severity.
However, motor score evaluation involves in-clinic visits with a trained medical professional, which are expensive and not always accessible.
Speech analysis provides an attractive avenue for tracking HD severity because speech is easy to collect remotely and provides insight into motor changes.
HD speech is typically characterized as having irregular articulation.
With this in mind, acoustic features that can capture vocal tract movement and articulatory coordination are particularly promising for characterizing motor symptom progression in HD.
In this paper, we present an experiment that uses Vocal Tract Coordination (VTC) features extracted from read speech to estimate a motor score.
When using an elastic-net regression model, we find that VTC features significantly outperform other acoustic features across varied-length audio segments, which highlights the effectiveness of these features for both short- and long-form reading tasks.
Lastly, we analyze the F-value scores of VTC features to visualize which channels are most related to motor score.
This work enables future research efforts to consider VTC features for acoustic analyses which target HD motor symptomatology tracking.

\end{abstract}

\section{Introduction}
Huntington Disease (HD) is a genetic, neurodegenerative disease that affects approximately 1 out of 10,000 individuals. Those afflicted with HD develop motor, cognitive, and psychiatric problems which worsen over time.
Although there is no known cure for HD, having the ability to track symptomatology is imperative in medical efforts for developing effective treatments.
Quantitative speech analysis provides a promising avenue for assisting medical professionals in characterizing symptom progression, since motor speech deficits are one of the most common symptoms observed in HD patients, occurring in roughly 90\% of cases~\cite{young1986huntington, skodda2014impaired, rusz2015automatic, rusz2013objective, vogel2012speech,diehl2019motor}.


Manifest HD can cause a variety of motor-related symptoms in speech resulting in reduced speech rate, abnormal speech rhythm, increased pauses, shorter vowel durations, higher intervocalization duration, and reduced articulation~\cite{skodda2014impaired, rusz2015automatic, rusz2013objective, vogel2012speech}.
These effects can also be compounded by co-morbidities, such as dysarthria (slurred speech) and/or bradykinesia (slow movement), which can impact speech production depending on the elicitation task. In premanifest HD, speech trends such as increased articulation rate, imprecise pace performance, and higher variability in syllable repetition are observed~\cite{skodda2016two}.
These clinical findings in manifest and premanifest HD suggest that speech can be a useful signal for disease progression tracking.
With this in mind, future applications should focus on automatic HD analyses that are robust and can handle a wide range of speech motor irregularities over HD severity.

Prior works have focused on differentiating between healthy control speakers and those with premanifest and manifest HD~\cite{perez2018classification,romana2020classification,riad2020vocal}. 
However, as we consider applications that can help medical professionals monitor symptom progression, it is important to consider severity on a continuous scale, such as predicting a motor score. There are several continuous scores useful for severity tracking; however, in this work we specifically focus on Total Motor Score (TMS) because it represents a critical component of the standard assessment used by medical professionals to evaluate HD severity~\cite{kieburtz2001unified}.

Clinical literature suggests that acoustic features which capture irregular articulatory movement in speech are ideal for HD motor tracking~\cite{skodda2014impaired}.
Vocal Tract Coordination (VTC) features are strong candidates for this task by quantifying articulatory coordination using correlations across low-level acoustic features. Although previous research has used these features to capture psychomotor articulation patterns in depression~\cite{williamson2014vocal,huang2020exploiting}, these features have not been explored at capturing the broad range of motor symptoms present in HD speech.

In this work, we investigate and analyze the effectiveness of VTC features for HD symptom progression tracking. We experiment with two extraction heuristics for VTC, eigendecomposition- (EVTC) and full-VTC (FVTC), in order to study the effect of dimensionality reduction on these features when characterizing motor symptomatology. We ultimately show that FVTC features achieve performance improvements over OpenSMILE, raw acoustic features, and EVTC when used for predicting TMS on read speech (i.e., the Grandfather Passage).
We also investigate how the length of the audio sample is related to downstream performance and analyze which FVTC feature channels are most relevant to motor score tracking.


\section{Related Work}

Prior work by Riad et. al. have shown that phonatory features are useful for classifying between control, premanifest, and manifest HD when using acoustic data from a max phonation task~\cite{riad2020vocal}. These phonatory features include jitter/shimmer, probability of voicing, and statistics over mel-cepstral features. The authors also show that these phonatory features achieve an R\textsuperscript{2} of 0.53 when predicting TMS. 
However, questions still remain as to whether these phonatory features can be extended to tasks that involve complex speech production (i.e., passage readings or spontaneous speech), which is the ideal for remote health monitoring applications. 
In our work, we extract similar phonatory and rhythm features using the OpenSMILE toolkit.


Previous work by Perez et. al. looked at capturing pauses, speech rate, and pronunciation from generated transcripts to automatically differentiate between healthy controls and gene positive HD participants who read the Grandfather Passage \cite{perez2018classification}. 
Their results show that neural networks could classify HD with up to 87\% accuracy and that the effectiveness of pauses, speech rate, and pronunciation were correlated with disease severity. 
However, a limitation of this approach is that the extracted features are directly tied to the performance of ASR systems, which typically underperform in unrestricted, disordered speech settings~\cite{christensen2012comparative,perez2020aphasic,fraser2013automatic}. In our current work, we remove this dependence on ASR systems for feature extraction by extracting acoustic features directly from audio.

Recent work by Romana et. al. looked at using vowel distortion to classify manifest vs. premanifest HD~\cite{romana2020classification}. The authors used manually extracted vowel segments from the Grandfather Passage. The authors demonstrate that these features are not only correlated to TMS but achieve 80\% classification accuracy of premanifest vs. manifest HD using trained models. Classification was performed at the speaker-level so features were computed over the reading passage. In contrast, rather than restricting to manually extracted vowels, the current work investigates acoustic features that can be extracted automatically over the entire passage.


\section{Data}

The dataset we use was collected at the University of Michigan~\cite{carlozzi2021understanding}. The dataset consists of 62 English-speaking participants (31 healthy, 31 with HD). Healthy control participants had no history of either neurological disorders or speech impairments. 
Individuals with HD had to have a positive gene test and/or clinical HD diagnosis by a physician. 

The motor severity of each participant was assessed using the TMS, the motor subsection of the Unified Huntington Disease Rating Scale (UHDRS)~\cite{kieburtz2001unified}, a standard assessment used by medical professionals to evaluate HD severity. The TMS is a holistic motor evaluation, which sums together individual motor subsections that target eye, trunk, gait, tongue, and speech movement. The TMS ranges from 0 (healthy) to 128 (severe). In our dataset, HD and Control speakers have an average TMS of 33.6 ($\pm$ 23.6) and 4.4 ($\pm$ 2.7) respectively.

HD participants were designated with premanifest HD if they had a diagnostic confidence rating of $<$4 on the last item of the TMS, while manifest HD was defined by a confidence rating of 4 on the last item of the TMS~\cite{liu2015motor}.
Within the manifest group, we label participants as early- or late-stage based on their Total Functional Capacity (TFC) scores, which range from 0 (low functioning) to 13 (high functioning)~\cite{shoulson1989assessment}.
Manifest early-stage was defined by a TFC score of 7-13, while manifest late-stage was defined by a TFC score of 0-6~\cite{marder2000rate}. The speaker breakdown according to HD severity is: 12 premanifest HD, 12 manifest early-stage, and 7 manifest late-stage.

The data include both interviews and a series of tasks~\cite{carlozzi2021understanding}.  All tasks were recorded using a table microphone at 44.1 kHz. This study focuses on the reading portion, where participants read the Grandfather Passage~\cite{reilly2012sherlock}. This reading task contains 129 words and 169 syllables and is commonly used to test for dysarthria in speech-language pathology.

\begin{table*}
\centering
\begin{tabular}{ lc|c|c|c|c|c|c } 
\toprule
 Features & RMSE & R\textsuperscript{2} & CCC & CCC-pre & CCC-early & CCC-late \\
 \hline
 Transcript Feats  & 18.0 (4.4) & 0.32 (0.29) & 0.52 (0.19) & 0.28 & 0.22 & 0.09 \\
 OpenSMILE IS10 & 20.3 (3.7) & 0.16 (0.34) & 0.39 (0.21) & 0.25 & 0.2 & 0.06 \\
 \midrule
 \multicolumn{3}{l}{\textit{Raw acoustic features}}\\
 \midrule
 \hspace{8pt} mfcc & 20.5 (4.0) & 0.16 (0.29) & 0.39 (0.19) & 0.08 & 0.12 & 0.07  \\
 \hspace{8pt} dmfcc  & 19.9 (3.9) & 0.21 (0.29) & 0.43 (0.17) & 0.21 & 0.04 & 0.17 \\
 \midrule
 \multicolumn{3}{l}{\textit{EVTC}}\\
 \midrule
 \hspace{8pt}mfcc  & 20.5 (4.1) & 0.17 (0.26) & 0.38 (0.18) & 0.18 & 0.09 & 0.19\\
 \hspace{8pt}dmfcc & 20.5 (4.0) & 0.13 (0.4) & 0.38 (0.21) & 0.24 & 0.21 & 0.24\\
 \hline
 \midrule
 \multicolumn{3}{l}{\textit{FVTC}}\\
 \midrule
 \hspace{8pt}mfcc  & 22.1 (3.1) & 0.04 (0.29) & 0.29 (0.19) & 0.2 & -0.2 & 0.12\\
 \hspace{8pt}dmfcc & \textbf{17.9 (3.5)} $\dagger$ & \textbf{0.32 (0.36)} & \textbf{0.51 (0.2)}$\dagger$ & \textbf{0.26} & \textbf{0.26} & \textbf{0.35}\\
 \hline
\end{tabular}
\vspace{4pt}
\caption{Elastic-net regression task when predicting Total Motor Score at the speaker-level using 10s segments. Results are averaged over 100 runs ($\pm$ std). $\dagger$ indicates that the marked performance is significantly higher than best baseline feature (raw-dmfcc). Significance is assessed at $p<0.05$ using the Tukey’s honest test on the ANOVA statistics.}
\vspace{-15pt}
\label{tab:TMS_regression}
\end{table*}

\section{Features}
\subsection{Vocal Tract Coordination Features}
Although previous works have shown success at using VTC-derived features to capture psychomotor articulatory differences when characterizing depression~\cite{huang2020exploiting,williamson2013vocal}, little work has been done on extending similar correlation-based features towards other domains. 
Clinical research has not only characterized HD speech as having irregular articulation but has even suggested that motor performance of the vocal tract and trunk extremities is controlled by the same mechanisms, presumably the basal ganglia~\cite{skodda2014impaired,diehl2019motor}.
This highlights the vocal tract as a key area of study for motor symptom tracking in HD and motivates the investigation of VTC features for characterizing the wide- range of articulatory symptoms in HD.


VTC features are derived from acoustic signals.
In our work we use librosa to extract 16-channel mel-frequency cepstral coefficients (mfcc), delta mel-frequency cepstral coefficients (dmfcc) using a hamming window of 25 ms with a 10 ms step size. We remove the 0th coefficient and apply speaker-level cepstral mean and variance normalization.



We calculate VTC features by applying auto- and cross-correlation functions to time-delayed channels (i.e., feature dimension) of a given acoustic signal, following the same approach as outlined in~\cite{huang2020exploiting}. For a given utterance, a normalized correlation $r_{i,j}^d$ is computed over the acoustic signal $x$ following equation~\ref{eq:corr}, where $i$ and $j$ are channels and $d$ is the time-delay used to shift a given channel.

\begin{equation}
\label{eq:corr}
r_{i,j}^d = \dfrac{\sum_{t=1}^{T-d}x_i[t]x_j[t+d]}  {\sqrt{r_{i,i}^0r_{j,j}^0}}
\end{equation}

We experiment with two feature formats for VTC. The first uses the full feature vector, which we call Full-VTC (FVTC). For FVTC, we concatenate all $r_{i,j}^d$, which results in a $N{\times}N{\times}D$ feature vector, where $N$ is the number of channels (15) across the entire utterance and a max delay of $D$ (80). The second format applies eigendecomposition across the channels of FVTC, which we call eigen-VTC (EVTC). EVTC was originally used in previous works~\cite{williamson2013vocal,williamson2014vocal} and results in a $N{\times}D$ feature vector. 
Our feature extraction code will be made publicly available~\footnote{https://github.com/matthewkperez/VTC-features}.

\subsection{Baseline Features}
Prior research has shown promise in using IS10 features from the OpenSMILE toolkit for emotion recognition, cognitive decline assessment, and characterizing disordered speech~\cite{pan2020acoustic,pan2020improving,an2015automatic}. We extract IS10 features which include low-level mfcc with appended deltas, shimmer, jitter, log-mel frequency band, and fundamental frequency features. Another benefit of IS10 is that it also contains some features which were used in ~\cite{riad2020vocal}. For each utterance, we compute broad statistics (mean, std, min, max, and median) over the frame-level feature matrix leading to a 352-dim feature vector.

For our second set of baseline features (raw acoustic features), we extract and use low-level acoustic features (mfcc and dmfcc) and compute broad statistics over the utterance in order to generate a fixed feature vector. We compute mean, std, min, max, and median following that of previous works~\cite{pan2020acoustic,alhanai2017spoken}.

\begin{figure}[t]
\centering
\includegraphics[width=1\columnwidth]{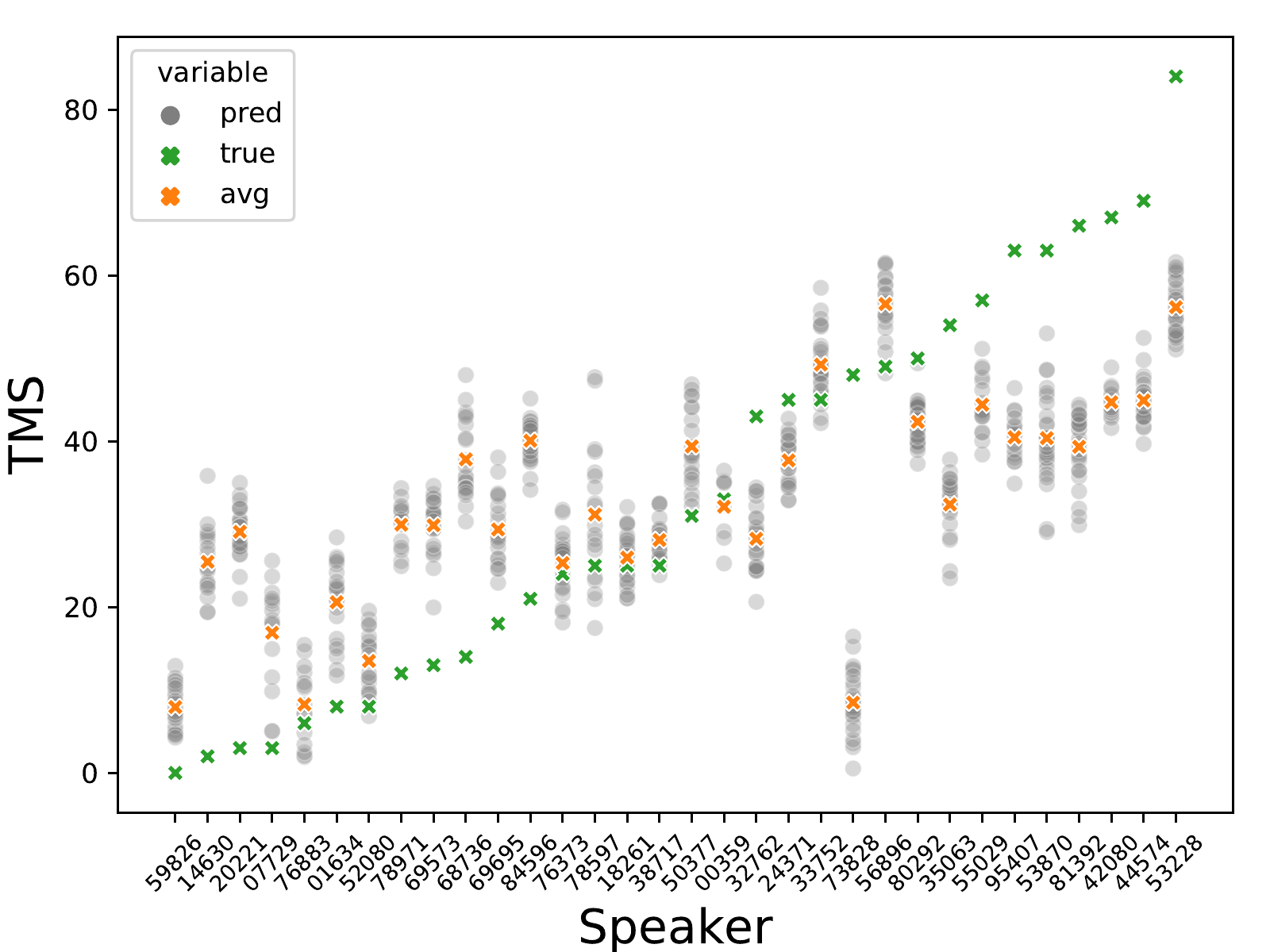}
\caption{Elastic-net model using FVTC-dmfcc features, where the top-75 features. 10s audio segments were used and the model was run a 100 times.}
\vspace{-5pt}
\label{fig:pred}
\end{figure}



\begin{figure*}[t]
\captionsetup[subfigure]{oneside, margin={0.5cm,0cm}}
\subfloat[Segment Variation: RMSE]{\label{fig:chunk_rmse}\includegraphics[width=50mm]{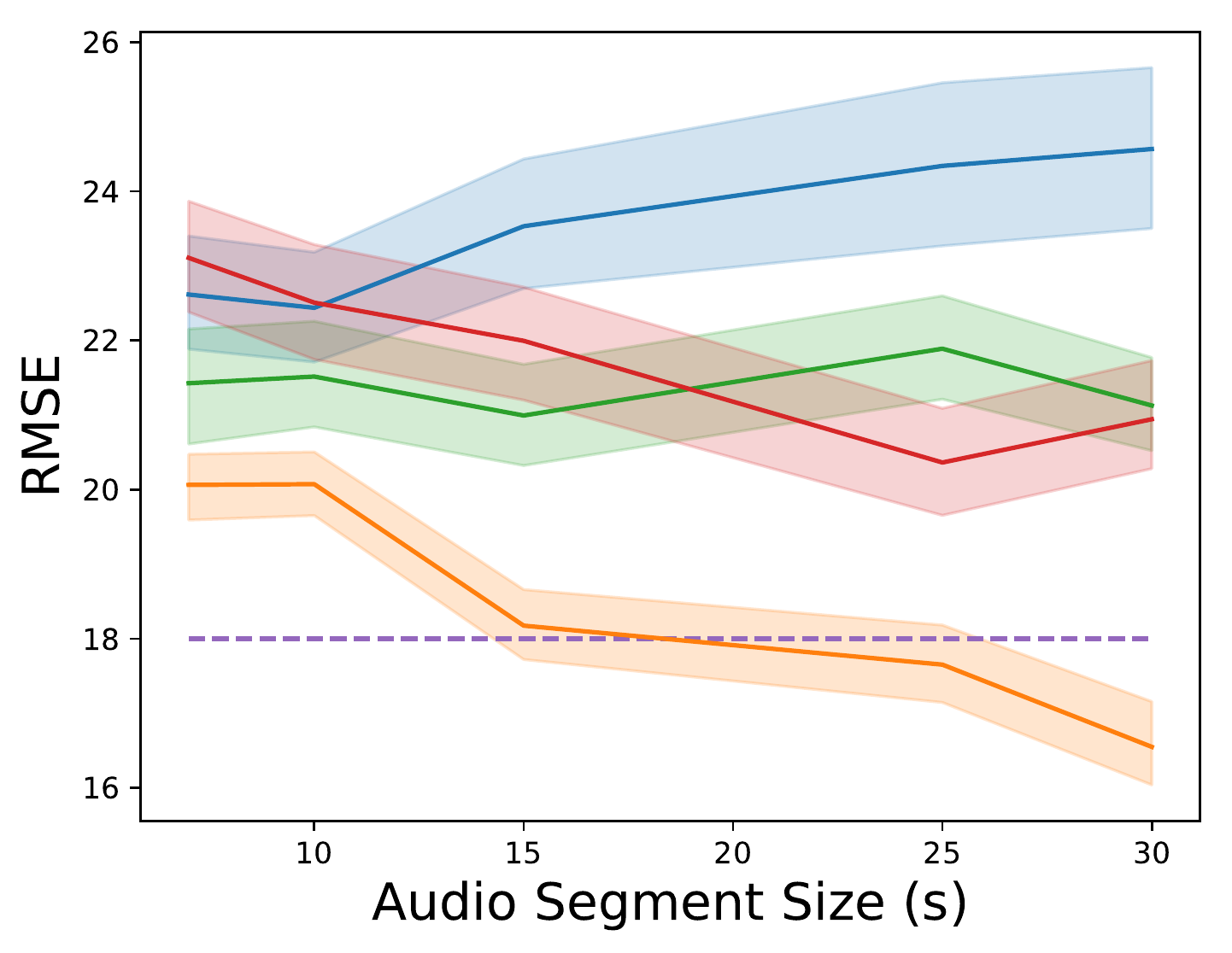}}
\subfloat[Segment Variation: R\textsuperscript{2}]{\label{fig:chunk_r2}\includegraphics[width=50mm]{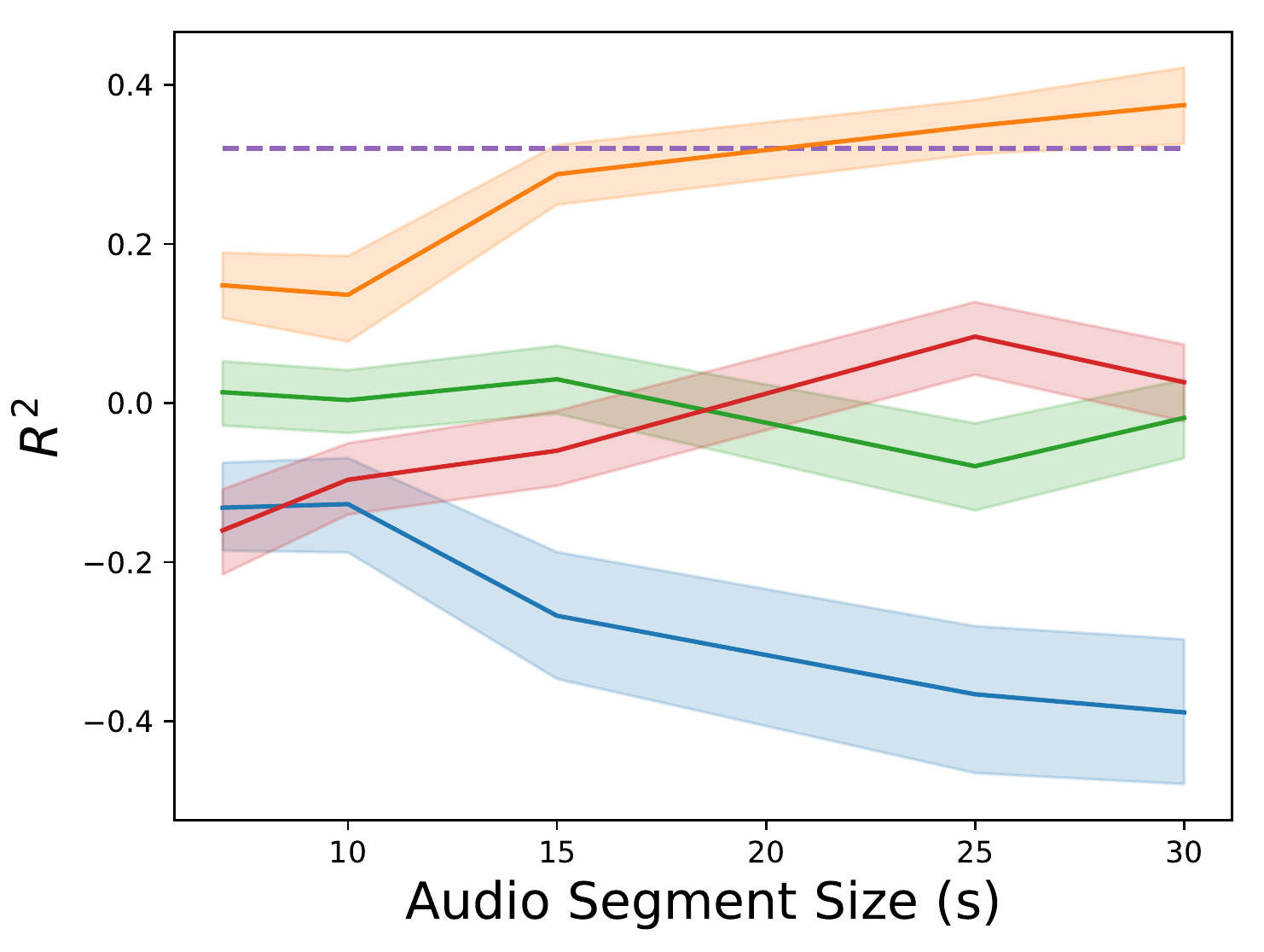}}
\captionsetup[subfigure]{oneside, margin={-1cm,0cm}}
\subfloat[Segment Variation: CCC]{\label{fig:chunk_ccc}\includegraphics[width=62mm]{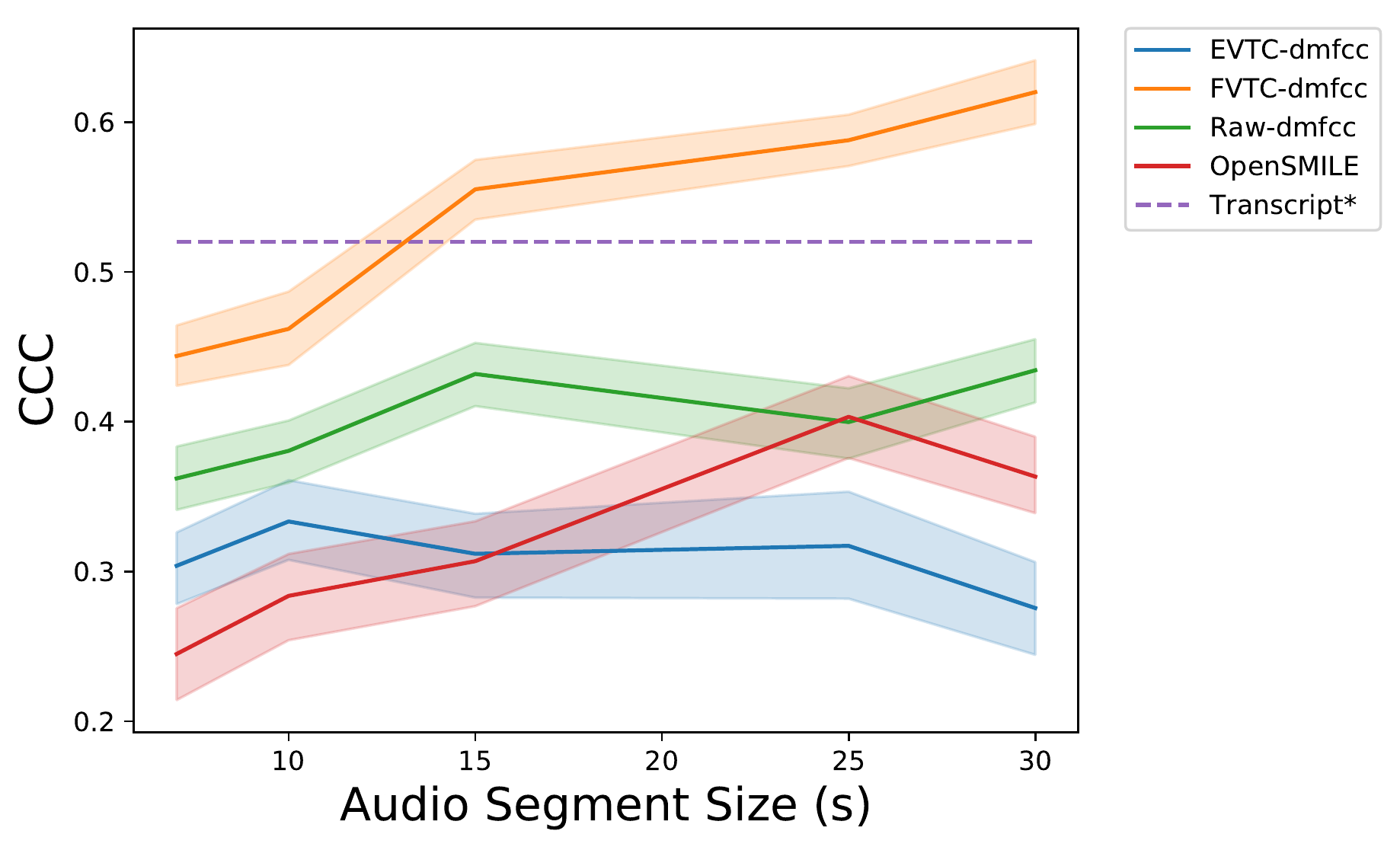}}
\caption{RMSE, R\textsuperscript{2}, and CCC using varying segment lengths to splice the Grandfather Passage. TMS estimation is performed at the segment-level.}
\vspace{-10pt}
\label{fig:chunk_TMS}
\end{figure*}


\begin{figure}[t]
\centering
\includegraphics[width=0.8\columnwidth]{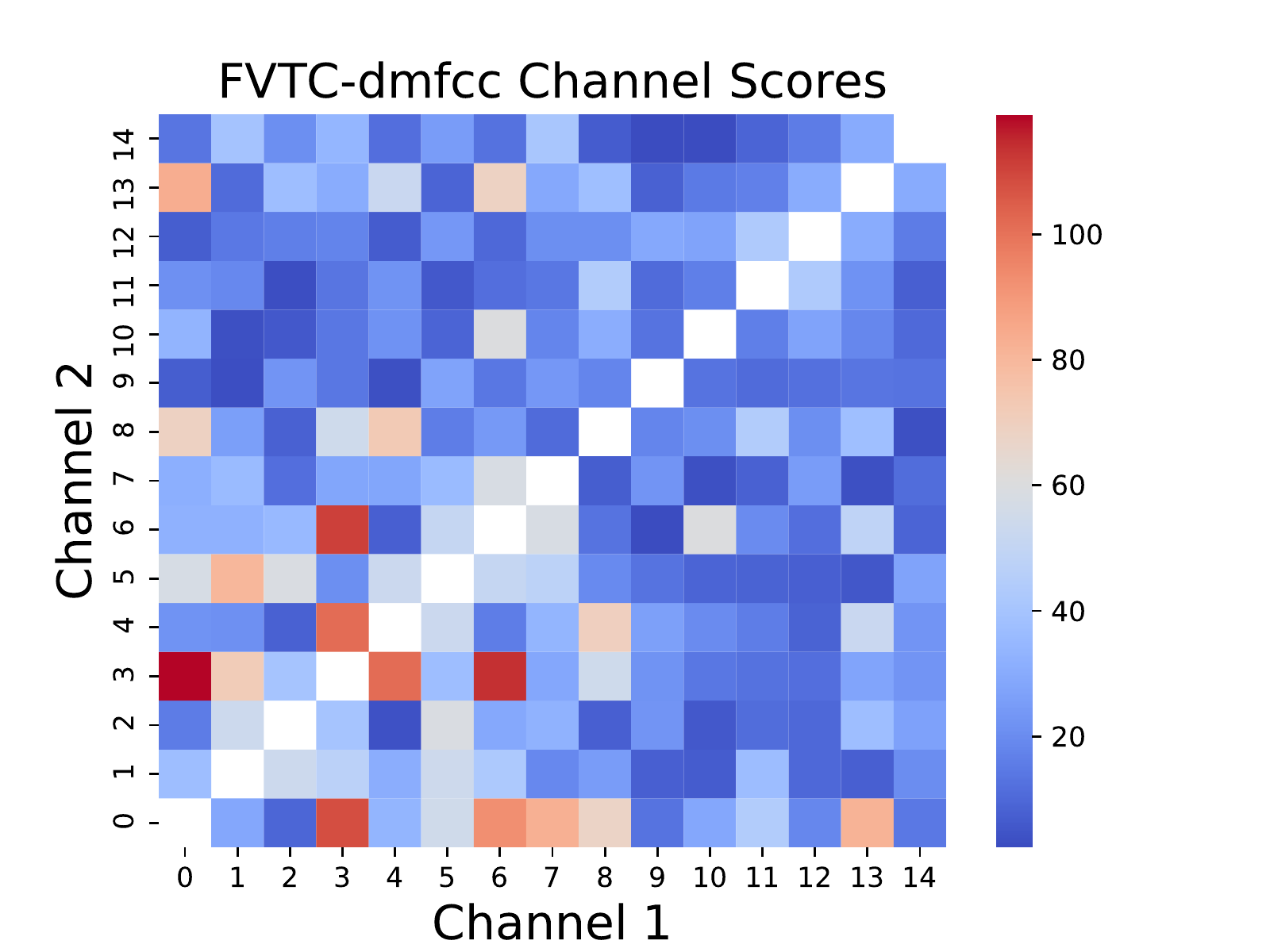}
\vspace{-8pt}
\caption{F-value scores for FVTC-dmfcc using 10s segments averaged over all 100 runs and channel maximum taken over all delays.}
\vspace{-12pt}
\label{fig:feat_analysis}
\end{figure}

\vspace{5pt}
\section{Experiment}
For all experiments, we initially segment the audio data into smaller chunks using a sliding window of 10s in order to increase the number of training samples in our dataset (Section~\ref{sec:TMS}) and then explore potential performance changes as a function of segment length (Section~\ref{sec:seg}). We assign the same TMS score to all segments from a given speaker.
We evaluate performance using a randomly selected 20\% held-out, speaker-independent test set. To account for randomness in test set selection, model initialization, and training, we run each experiment $100$ times and present metric averages ($\pm1$ std) over all runs. 
We downsample the control population to a randomly selected seven speakers due to the large TMS imbalance (i.e. majority of speakers are control or premanifest). This selection was motivated by the number of late-stage HD speakers (seven) and serves to ultimately circumvent biasing the classifier towards low TMS scores.

We perform feature normalization and select the top 75-features (the upper bound for our raw acoustic feature set) based on F-value scores using the training set. Selected features are then used to train an elastic-net model, where L1 and L2 regularization is combined using default hyperparameters, (C=1 and ratio=0.5)~\cite{pedregosa2011scikit}. Elastic-net is utilized for our TMS regression task due to its ability to handle multicollinearity and perform variable selection and regularization simultaneously~\cite{riad2020vocal,alhanai2017spoken}.

\subsection{TMS Score Prediction}
\label{sec:TMS}
We use the elastic-net model to predict TMS for each segment and perform speaker-level averaging over all segments for the final TMS prediction.
We evaluate system performance using root-mean-square error (RMSE), R\textsuperscript{2}, and concordance correlation coefficient (CCC) between the predicted and target scores. In addition, we compute the CCC for each HD severity (premanifest, early, and late) over all runs. With this, we can better understand what types of speakers the model is able to fit with regards to HD severity. To assess the statistical significance we perform a Tukey honest significance test using $p<0.05$.

Our goal is to identify how well VTC features (FVTC and EVTC) characterize TMS in comparison to other baseline feature sets. Our results presented in Table~\ref{tab:TMS_regression} shows that FVTC-dmfcc outperforms all baseline methods across all evaluation metrics when estimating TMS. We note significant improvements in RMSE and CCC, achieving average values of 17.9 and 0.51 respectively. We also note that when comparing EVTC and FVTC, eigendecomposition seems to benefit mfcc; however, for dmfcc using FVTC features provides better performance. Ultimately, these results highlight FVTC-dmfcc as a promising acoustic biomarker for modeling motor symptomatology.

Figure~\ref{fig:pred} illustrates the performance of FVTC-dmfcc for all speakers by showing the ground truth (green), speaker prediction (grey), and mean prediction (orange) TMS. One point of emphasis is the model's performance for higher TMS, where we can see a stronger correlation (CCC = 0.35) for late-stage speakers, compared to premanifest- and early-stage speakers (Table~\ref{tab:TMS_regression}). One outlier to note is speaker 73828, which the model incorrectly estimates as having low TMS. Although further investigation is required, initial analyses show that this speaker is an outlier across all feature sets, which seems to suggest that this speaker's motor issues may not manifest in speech. Additionally, it's important to note that Transcript Feats were extracted over the entire passage rather than shorter intervals (i.e. 10s in Table~\ref{tab:TMS_regression}). We further explore the relationship of audio length on the impact of the baseline and VTC features in the next section.

\subsection{Segment Size Variation}
\label{sec:seg}
Our next experiment analyzes how the segment-level performance of our system changes as the length of the segments is varied.  In the previous analysis, we restricted segments to 10s due to the relatively small dataset size. However, in this section, we analyze potential performance increases resulting from longer segment lengths to inform future studies.  We segment the grandfather passage into 7, 10, 15, 20, 25, and 30 second segments and perform TMS estimation at the segment-level.

Figure~\ref{fig:chunk_TMS} shows the performance of RMSE, R\textsuperscript{2}, and CCC for our different feature sets when varying audio segment size. We can see performance improvement in RMSE, R\textsuperscript{2}, and CCC when using FVTC-dmfcc across all segment sizes. 
OpenSMILE, raw feats, and FVTC all show improvements when using longer audio segments. We hypothesize that for OpenSMILE and raw feats this is likely due to taking broad statistics over longer audio sequences. 
In the case of FVTC, we demonstrate that longer audio intervals allow for more robust and generalizable correlation computation, which is a finding that is also consistent with previous work~\cite{huang2020exploiting}. We also see that for longer audio lengths FVTC starts to outperform Transcript Feats (which utilize the full passage length).


These results motivate the need for longer form passage readings since nearly all acoustic features seem to benefit from longer audio. However, in the case of both short- and long-form audio readings, FVTC-dmfcc should be an acoustic feature considered for characterizing TMS.




\subsection{FVTC Channel Analysis}
Our last experiment analyzes the FVTC-dmfcc feature space using the scores derived from F-value between feature and TMS. Figure~\ref{fig:feat_analysis} shows a heatmap where the x- and y-axis represent the $i$ and $j$ channels used for the FVTC-dmfcc features. Feature scores are averaged over all 100 iterations and the maximum score is taken for each channel across all delays. Channels 0-6 seem to represent a dense amount of relevant information for TMS regression, accounting for 66\% of the final selected features, which suggests that correlations across channels in the low frequency space are related to TMS. 




\section{Conclusion}
In this work we present a novel investigation into the effectiveness of Vocal Tract Coordination (VTC) features for characterizing motor symptomatology in HD speech. We show that FVTC-dmfcc significantly outperforms all other acoustic features in predicting TMS with regards to RMSE and CCC by achieving average values of 17.9, 0.32, and 0.51 respectively. 
Furthermore, we demonstrate that FVTC-dmfcc outperforms other acoustic features when classifying varied length segments, which suggests these features should be considered when analyzing both short- and long-form reading tasks. We also show that FVTC-dmfcc performance improves as audio length increases, which motivates data collection methods to push for long form audio readings. Lastly, we analyze channel importance by plotting F-value scores and demonstrate that low-frequency channels are most relevant to TMS.

Future work, will encompass further analysis into VTC features and their ability to characterize disordered speech. We plan to investigate other disorders, such as Parkinson's Disease. Additionally, as we start to consider passive, remote health monitoring applications we would like to relax the dependence on read speech tasks and utilize free speech as audio input.

\section{Acknowledgements}
The authors would like to thank Thomas Quatieri, James Williamson, and Zhaocheng Huang for their helpful discussions.
This material is based in part upon work supported by the NSF-GRFP, NIH, National Center for Advancing Translational Sciences, National Institute of Neurological Disorders and Stroke. Any opinions, findings, and conclusions or recommendations expressed in this material are those of the authors and do not necessarily reflect the views of the funding sources listed above.

\bibliographystyle{IEEEtran}
\bibliography{main}
\end{document}